\documentclass[12pt]{article}
\begin{document}
\begin{center}
{\large  On the $\pi\pi$ - scattering lengths in the theory with effective Lagrangian}

{\bf   E.P.  Shabalin \footnote{e-mail: shabalin@itep.ru}} \\
Institute of Theoretical and Experimental Physics; National
Research Center Kurchatov Institute, Bol'shaya Cheremushkinskaya
ul., 25,  Moscow, 117218 Russia.
\end{center}
\begin{abstract}
Use of the effective Lagrangian incorporating both the scalar and pseudoscalar mesons gives a possibility
to calculate the  $\pi\pi$-scattering lengths without attraction of the ChPT theory.
\end{abstract}

\section{Introduction}
The question about  the numerical quantities of the  $\pi\pi$ -scattering lengths engages the minds of theoreticians and the experimentalists so far.  Nowdays, a number of the observed events of $\pi\pi$ - scattering  increased many times, that gives a
possibility to verify the different approaches to work out a theory of the low-energy $\pi\pi$ - interaction.
It is clear, that an approach based on the use of the Lagrangian scheme looks as more convenient  for deep understanding
of the theorems concerning the soft pions.
Just in this scheme, it is possible to obtain the precise results of the current algebra even on the  "trees" level, that is, on the level of the diagrams without loops [1].

In this paper, this peculiarity of the Lagrangian scheme will be used for the calculation of the   $\pi\pi$ scattering lengths.  And besides, the  conformity of the properties of the QCD objects and objects of the real world will be  ensured also.
In QCD, the spinless flavoured objects are $\bar q_R t^a q_L$ and their Hermitian conjugate. These objects
are formed by opposite-parity components. For this property to be reproduced in the Lagrangian of real particles, it must be expressed in terms  of the matrix
\begin{equation}
U = (\sigma_a + i\pi_a)t_a,
\label{1}
\end{equation}
where $t_0 = \sqrt{1/3} $ ,  $t_{1....8} = \sqrt{1/2}_{1,....8}$ and $\sigma_a$ and  $\pi_a$ are nonets of,
respectively, scalar and   pseudoscalar mesons.
This idea is not new and was used in $ [2] $ and $ [3 ]$, but we go over directly to the final form of the Lagrangian containing
all forms of breakdown of chiral symmetry [4], [5]:
\begin{equation}
\begin{array}{ll}
L = \frac{1}{2}Tr(\partial U_{\mu}\partial U^+ _{\mu})  - cTr(UU^+
- A^2t^2_0)^2 - c\xi( Tr(UU^+ - A^2t^2_0))^2 + \nonumber \\
\frac{F_{\pi}}{2\sqrt{2}}Tr{(M(U +U^+)} + \Delta L^{U(1)}_{PS}.
\end{array}
\label{2}
\end{equation}
These forms permit to express all properties of $\sigma$ - mesons using the properties of $\pi$ - mesons
and the parameters $R=F_K/F_{\pi}$ and $\xi$. The values of them can be found from the data on the decays
$K,\pi \to \mu \nu$ and identification of the scalar $\sigma_{\pi}$  - meson with the resonance $a_0(980)$  [4], [5], [6].

In a theory with the broken $U(3)_L\otimes U(3)_R$ chiral symmetry there are two isosinglet $\sigma$-particles having nonzero vacuum expectation values $<\sigma>$. As a result,  the Lagrangian(2) contains
the vertices $<\pi\pi|\sigma>$. Then, the set of the pole diagrams with the intermediate $\sigma$ - meson appears. The amplitude of the $\pi\pi \to \pi\pi$ - scattering acquires the form:
\begin{equation}
\begin{array}{ll}
T_{\sigma} = <\pi_k(p'_1)\pi_l(p'_2)|\pi_i(p_1)\pi_j(p_2)>  =
A_{\sigma}\delta_{ij}\delta_{kl} + B_{\sigma} \delta_{ik}\delta_{jl} + C_{\sigma} \delta_{il}\delta_{jk},
\label{3}
\end {array}
\end{equation}
where
\begin{equation}
\begin{array}{ll|}
A_{\sigma} = (s-\mu^2)\sum_{n=1,2} \frac{G_n}{m^2_{\sigma_n}-s}, \quad
 B_{\sigma} = (t-\mu^2 )\sum_{n=1,2}
 \frac{G_n}{m^2_{\sigma_n}-t},\nonumber \\
 ~~ \nonumber \\
 C_{\sigma} = (u-\mu^2)\sum_{n=1,2} \frac{G_n}{m^2_{\sigma_n} - u},    \quad \mu \equiv m_{\pi},
\end{array}
\label{4}
\end{equation}
and where
\begin{equation}
\begin{array}{ll}
s = (p_1 +  p_2)^2, \quad t = (p_1 - p'_1)^2, \quad u = (p_1 - p'_2)^2, \quad
 G_n=\frac{g^2_n}{m^2_{\sigma_n} - \mu^2}.
\end{array}
\label{5}
\end{equation}
In the theory specified by the Lagrangian (2) , the following relation holds:
\begin{equation}
\frac{G_1}{m^2_{\sigma_1} - \mu^2} +\frac{G_2}{m^2_{\sigma_2} - \mu^2}=\frac{1}{F^2_{\pi} }, \quad
F_{\pi} = 93 \:{ MeV}.
\label{6}
\end{equation}
For the case of fixed total isospin of a system of initial pions, the expressions for the amplitudes are given in [7]:
\begin{equation}
T^{(0)} = 3A + B + C.
\label{7}
\end{equation}
\begin{equation}
T^{(1)} =B-C
\label{8}
\end {equation}
\begin{equation}
T^{(2)} = B +C.
\label{9}
\end{equation}
The decomposition of the isotopic amplitudes into amplitudes corresponding to fixed values of the orbital angular momentum is given by
\begin{equation}
T^{(I)} = 32\pi\sum_{l=0}^{\infty} (2l+1)t^{(I)}_l(s)P_l(\cos \theta).
\label{10}
\end{equation}
It follows from (10) that the partial-wave amplituda $t^{(I)}$ is:
\begin{equation}
t^{(I)}_l(s) = \frac{1}{64\pi} \int_{-1}^{1} T^{(I)}P_I(\cos\theta) d \cos\theta.
\label{11}
\end{equation}
The scattering lengths arise from the expansion
\begin{equation}
t^I_l(s)m^{-1}_{\pi} = q^{2l}[a^{(I)}_l + b^{(I)}_l q^2 + O(q^4)], \qquad q^2=\frac{s}{4}-\mu^2.
\label{12}
\end{equation}
To calculate $a^{(0)}_0$ and $a^{(2)}_0$, we make use of the relations (3,7,9,11). The scattering length
$a^{(1)}_1$ will be considered later, since besides the $\sigma$ - mesons, the $\rho$-mesons also
contribute into this scattering length. We begin calculations from $a^{(2)_0}$, because this scattering
length,
according to (9) and (7),enters into $a^{(0)}_0$.

\section{The scattering length $a^{(2)}_0$}
According to (9)
\begin{equation}
T^{(2)}_{\sigma} = \sum_{n=1,2} G_n\left(\frac{t - \mu^2}{m^2_{\sigma_n} - t} +\frac{u - \mu^2}{m^2_{\sigma_n}
-u} \right).
\label{13}
\end{equation}
On the threshold $t$ and $u$ are equal to zero. Using the expressions for masses and coupling constants
of the  $\sigma_n$ - mesons [5] and the last numerical data on their values [6], we obtain:
\begin{equation}
T^{(2)}_{\sigma} = - \left(\frac{2G_1\mu^2}{m^2_{\sigma_1|}} +\frac{2G_2\mu^2}{m^2_{\sigma_2}}  \right) =
- 4.3171.
\label{14}
\end{equation}
According to (11) and (12)
\begin{equation}
a^{(2)}_0  = - 0.04294m^{-1}_{\pi}.
\label{15}
\end{equation}
This result is in agreement with the result of analysis of the $K^{\pm} \to \pi^{+}\pi^{-}e^{\pm}\nu$ decay,
based on the statistics of 1.13 million decays [8]:
\begin{equation}
a^{(2)}_0 = (- 0.0432 \pm 0.0086_{stat} \pm 0.0034_{syst} \pm 0.0028_{th})m^{-1}_{\pi}
\label{16}
\end{equation}

\section{ The scattering length $a^{(0)}_0$}

In accordance with (7) , the amplitude of $S$-wave with the isospin 0 looks as
\begin{equation}
T^{(0)} = \sum_{n=1,2} G_n \left( 3\frac{s - \mu^2}{m^2_{\sigma_n} - s} \right) +T^{(2)}_{\sigma}.
\label{17}
\end{equation}
At the threshold, the first addendum in (17) performs into
\begin{equation}
9\mu^2 \sum_{n=1,2} G_n(m^2_{\sigma_n} - 4\mu^2)^{-1} = 23.3335.
\label{18}
\end{equation}
Adding the result (14) for $T^{(2)}_{\sigma}$, we come to
\begin{equation}
T^{(0)}_{\sigma} = 19.0164.
\label{19}
\end{equation}
And the final result for $a^{(0)}_0$ is:
\begin{equation}
(a^{(0)}_0)_{\sigma} = \frac{19.0164}{32\pi m_{\pi}} = 0.18916m^{-1}_{\pi}.
\label{20}
\end{equation}
This value agrees with the experimental one:
\begin{equation}
(a^{(0)}_0)_{exp} = (0.197\pm 0.010)m^{-1}_{\pi}
\label{21}
\end{equation}
obtained from the analysis of all data near  threshold of the reaction $\pi N  \to  \pi\pi N$ \quad [9].

The data obtained for $a^{(0)} _0$ by the collaboration E865 [10] depend from the  Models used at analysis.
In particular, in the  Model A  $a^{(0)}_0 = (0.184 \pm 0.010)m^{-1}_{\pi}$,  in the Model B  $a^{(0)}_0=
(0.179 \pm 0.033)m^{-1}_{\pi}$ and in the Model C  $a^{(0)}_0=(0.213 \pm 0.013)m^{(-1)}_{\pi}$. \footnote{See the Table 6 in [8]}
However, these results appear after taking into account the isospin corrections, also depending from the
Models.

Our  result (20) does not demand any additional model corrections.

\section{The scattering length $a^{(1)}_1$}
In the used by us theory,  besides the $\sigma$ - mesons, the intermediate $\rho$ - mesons also give
a contribution into $a^{(1)}_1$. In the present paper, a nature of the $\rho$ - mesons will be not associated with the vector quark current, as it was  considered usually, but their nature will be associated with the
divergence of vector quark current [11 - 15].

A contibution of the $\sigma$ - mesons into isovector amplitude is:
\begin{equation}
T^{(1)}_{\sigma} = \sum_{n=1,2}G_n \left[\frac{t - \mu^2}{m^2_{\sigma_n} - t} - \frac{u - \mu^2}{m^2_{\sigma_n} -u} \right].
\label{22}
\end{equation}
According to the relations (11) and (12), the part of scattering length  $a^{(1)}_1$ produced by $\sigma$ -
mesons is:
\begin{equation}
 (a^{(1)}_1)_{\sigma} = \frac{1}{24\pi m_{\pi}} \left[ \frac{g^2_1}{m^4_{\sigma_1}} + \frac{g^2_2}{m^4_{\sigma_2}} \right].
\label{23}
\end{equation}
Using the numerical values of $m_{\sigma_{1,2}}$ and $g_{1,2}$, we find:
\begin{equation}
(a^{(1)}_1)_{\sigma} = 0.02744m^{(-3)}_{\pi}.
\label{24}
\end{equation}

The part of the isovector amplitude produced by the intermediate $\rho$ - mesons looks like the relation
(3), but with the differnt A, B and C [15]. Namely,
\begin{equation}
A_{\rho} = \frac{1}{M^2_{\rho}} \left(g^2(u)\frac{(s-t)u}{M^2_{\rho} -u} + g^2(t)\frac{(s-u)t}{M^2_{\rho} -t} \right).
\label{25}
\end{equation}
\begin{equation}
B_{\rho} = \frac{1}{M^2_{\rho}} \left(g^2(s)\frac{(t-u )s}{M^2_{\rho} -s} + g^2(u)\frac{(t-s)u}{M^2_{\rho} -u} \right).
\label{26}
\end{equation}
\begin{equation}
C_{\rho} = \frac{1}{M^2_{\rho}} \left(g^2(s)\frac{(u-t)s}{M^2_{\rho} -s} + g^2(t)\frac{(u-s)t}{M^2_{\rho} -t} \right).
\label{27}
\end{equation}
In accordance with (8)
\begin{equation}
T^{(1)}_{\rho} = B_{\rho} - C_{\rho}.
\label{28}
\end{equation}
As the explicit form of $g(x=s,t,u)$ is not specified by our theory, we are forced to resort to the
phenomenological relation elaborated in [15]:
\begin{equation}
g(x) = g_{\rho} exp \left(0.7855 \left[\frac{x}{2M^2_{\rho}} - \left(\frac{x}{2m^2_{\rho}}\right)^2 \right] \right),
\qquad x \le M^2_{\rho}.
\label{29}
\end{equation}
As we are interesting in behavior of the partial wave $t^{(1)}_1(s)$ near threshold, we obtain the following result:
\begin{equation}
(t^{(1)}_1)^{threshold}_{\rho} = \frac{4\mu^2}{3\pi M^2_{\rho}} \left(\frac{2g^2(4\mu^2)}{M^2_{\rho} - 4\mu^2} +\frac{g^2(0)}{M^2_{\rho}} \right)
\label{30}
\end{equation}
Using the experimental value of $g(M^2_{\rho})$=5.9764 and the formula (29), we find:
\begin{equation}
g(0) = 4.9108, \qquad g(4\mu^2) =5.5520.
\label{31}
\end{equation}
The part of the scatterinag length $a^{(1)}_1$ produced by the intermediate $\rho$ - mesons turns out to be:
\begin{equation}
(a^{(1)}_1)_{\rho} = 0.005918m^{(-3)}_{\pi}.
\label{32}
\end{equation}
Together with the part (24) produced by the $\sigma$ -mesons we get:
\begin{equation}
(a^{(1)}_1)^{total} =0.03336m^{-3}_{\pi}.
\label{33}
\end{equation}

\section{Conclusion}

The requirement of conformity between the properties of  the QCD objects and objects of the real world gives rise to necessity of existance of the scalar mesons, that, as it turned out, play the principal role in the low-energy $\pi\pi$ - interactions.
Being the chiral partners of the pseudoscalar mesons, the scalar mesons possess quite definite properties, ascribed by the structure of the Lagrangian (2). Their masses and coupling constants depend only on
two parameters:$ R = F_{K}/F_{\pi}$  and  $\xi$, the values of which are determined experimentally.
The found by us scattering lengths are:
\begin{equation}
a^{(0)}_0=0.18916m^{-1}_{\pi}, \; a^{(2)}_0=-0.04294m^{-1}_{\pi}, \;a^{(1)}_1=0.03336m^{(-3)}_{\pi}.
\label{33}
\end{equation}
And they did not require to attract the special models, concerning these scattering lengths.
In our theory, the current-algebra prediction [16]:
\begin{equation}
\frac{2a^{(0)}_0 - 5a^{(2)}_0}{18\mu^2{\pi}a^{(1)}_1} = 1
\label{34}
\end{equation}
is satisfied to within 1.24\%.

\end{document}